December 11, 2018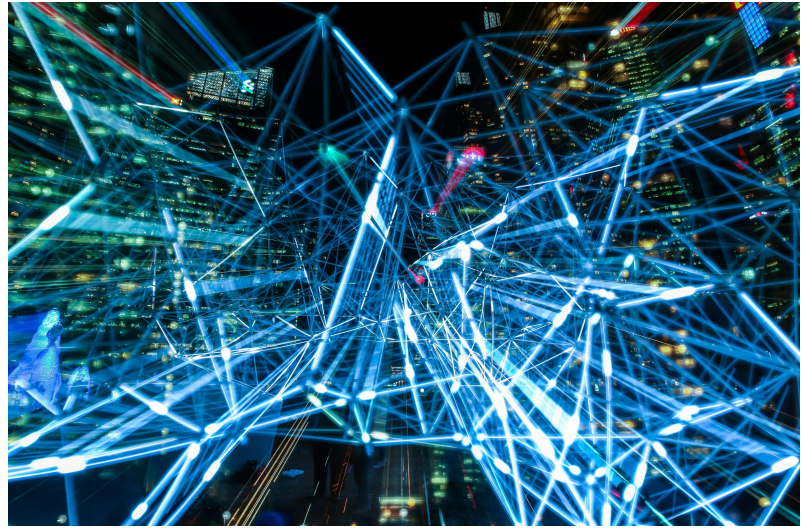

# Analysis and Consideration on Management of Encrypted Traffic

**White paper by the EU-H2020 MAMI project (grant agreement No 688421)**

Pedro A. Aranda (UC3M), Diego López (Telefonica), Thomas Fossati (Nokia)

The Measurement and Architecture for a Middlebox'ed Internet (MAMI) project is an EU-H2020 funded project that started on Junauray 2016. One of the focus areas of MAMI is the development and investigations of architectural approaches to enable explicit middlebox coordination to support traffic encryption. This paper is summarizing some of the finding for an audience from the industry. Further information about the project as a whole can be found on the [MAMI project website](#).Encrypted traffic poses new and unique challenges for network operators because information that is useful or necessary for management purposes is not accessible anymore. This paper examines proposed approaches to provide end-to-end encryption as well as ways to provide the access to the needed information for network management. The two main approaches we consider are

1. extending and adapting a widely deployed protocol such as TLS, so that information necessary for network management can be obtained; and
2. introducing a new protocol layer, such as PLUS, that contains the necessary information outside of the encrypted payload.

In this paper we discuss different aspects of these approaches in order to give recommendations for future work and suggest to raise awareness by establishing venues where discussions between interested parties can take place.

The original trust model, adopted by the Internet, considered it acceptable to use clear text communications for end-to-end communications. However, the need for communications security, at least in parts of the end-to-end communication, arose quickly. As the Internet evolved to support sensitive day-to-day operations (i.e. banking, private communications, etc.), communications security was the only means to protect the end-to-end communication from eavesdropping (sniffing) or modification (Man-in- the-middle (MITM) attacks). This protection is typically achieved by using cryptographic protocols like TLS [1, 2, 3], or the use of encrypted tunnels. The concerns are greater when radio transmission can be easily intercepted. For example, the impact of the everyday scenario using WiFi networks is reasoned in [4]. In addition, RFC 7258 concludes a consensus in the Internet Engineering Task Force (IETF), that the pervasive use of monitoring of Internet traffic has to be considered an attack [5].

**Evolution of Encryption Technologies on the Internet**

During its evolution, and in order to cover specific needs, intermediate devices that inspect or even modify end-to-end traffic have been introduced and accepted in the Internet. Such intermediate nodes interacting with the end-to-end traffic are known as middleboxes. One example middlebox is the Network Address Translator (NAT) defined in RFC1631 [6] needed to overcome the depletion of the IPv4 address space until IPv6 is fully deployed. Another example is Network Address Translation-Protocol Translation (NAT-PT) [7] as a transition mechanism to accelerate IPv6 deployment. Too guarantee that end-to- end encryption is compatible with these mechanisms, the IETF later produced compatibility requirement documents [8, 9].

Since middleboxes have established themselves as a mean to provide in-network services, there has been extensive research on how they can interact in an orderly manner with endpoints in end-to-end encrypted communications and different protocols have been developed to implement this interaction. While middlebox interaction is still a debated topic in the IETF, European Telecommunications Standards Institute (ETSI) has started work on standardising it.

The MAMI project organised the MAMI Management and Measurement Summit (M3S) that examined the challenges of pervasive encryption on network management in the Internet. The workshop report [10] concludes that despite encryption, independent measurability of the Internet should be provided and that the trend should be to introduce middlebox transparency in order to replace transparent middleboxes.

# Impact of Encryption on Network Management

The impact of pervasive encryption on network operations is discussed in [11]. This notes that there are legitimate situations where decryption is needed. One such situation is mitigation of Denial-of-Service (DoS) attacks, where it is desirable to protect and favour legitimate traffic and block illegitimate traffic. Another is the use of admission control techniques to Differentiated Services [12, 13]-enabled networks as a way to map traffic flows to network treatment most favourable to them (for example, optimised treatment of real-time traffic [14]). [15] examines the impact of encryption on network operations and points out that there will always be trade-offs between what parts of the communication are encrypted and the ability to operate and maintain the Internet as well as to develop transport protocols to keep up with the technological evolution. In addition, as noted in [11], there is a mandate for operators to implement lawful intercept to assist agencies in crime prevention. Finally, there are scenarios, such as providing Internet access in rural areas using satellite access, where standard TCP would achieve very low throughput. In this case, highly invasive techniques are used such as TCP transparent proxies or asymmetric TCP acceleration can have significant benefit. These techniques are not compatible with end-to-end encryption [16].

Future evolutions where access control techniques or the lack of them will be critical include Deterministic Networking [17]. In the absence of techniques that enable identify special traffic classes in the Internet (including the use of specialised middleboxes), the use cases currently considered by the Working Group (WG) [18] will lead to the creation of parallel, isolated Internets to support them.

### Impact on Mobile Networks

The GSMA, through its Internet Working Group, has considered the implications of a new Internet stack in the current network management practices of mobile network operators. Two white papers have been published, analyzing the specific requirements of radio access, the current techniques in use to address these requirements, and the impact of the Internet stack evolution on these techniques. The documents recommend further investigation, with the goals of assessing the technologies analyzed and to contribute towards standardizing solutions. In addition, a workshop between the GSMA and the Internet Architecture Board (IAB) help in September, 2105 discussed the implications of bandwidth optimisation tools for mobile networks and pervasive encryption [19]. [20] describes approaches allowing operators to manage network data that is increasingly encrypted, and delivered via new protocols and congestion-control algorithms. It also covers non-technical considerations, which include regulation and security. The document provides a set of recommendations for the coming 5G era, where there is expected to be a radical increase in the range of traffic types, and overall load placed on the network. Taking these approaches as a starting point, MAMI has conducted a set of experiments to demonstrate the usefulness of providing information about a flow crossing a 5G networks [21] that is either loss or latency tolerant.

Once middlebox interaction is assumed to take place, another question arises: how are the the end- points to be notified that cooperating middleboxes operate on the path through a domain. This can be accomplished by either out-of-band configuration, maintaining Network Operator information base that would need to be passed to the endpoints by device configuration protocols such as Dynamic Host Con- figuration Protocol (DHCP) or by another mean, e.g. Domain Name Service (DNS). Another alternative explored by developers is to provide in-band device discovery, i.e. where a cooperating middlebox signals its presence and capabilities during the handshake phase (e.g., e [22]).

# Analysis of Middlebox Cooperation Mechanisms

Depending on the mindset, middlebox cooperation can be approached from an evolutionary point of view in different ways, i.e. whether to upgrade existing protocols, whether to add new features, caring or not about backwards compatibility, or the addition of new layers to the protocol stack that can carry the additional information that can be utilised to support middlebox cooperation.

# TLS-based Approaches

TLS provides encrypted end-to-end communications, and does not not allow for in-path devices to interact with the traffic. Acknowledging that privacy of communications is important and should be preserved, a number of in-network services have evolved during the last years, allowing, e.g. for traffic optimisation, allowing Internet connectivity to expand to regions where network deployment is challenging or enhancing the end users' QoE, traffic monitoring for parental protection, etc. These services are normally provided by Network Operators using middleboxes. In this section, we look at different, proposed extensions or additions to TLS to are designed to enable middleboxes cooperation.

- ## Multi-context TLS

Multi-context TLS (mcTLS) [23, 24] extends version 1.2 of TLS. It allows endpoints to identify trusted middleboxes that will interact with the traffic. It is based on the following principles: 1. both endpoints need to explicitly approve each middlebox; 2. each middlebox only interacts with the parts of the traffic it needs to do its job; 3. the identity of each middlebox can be verified by the endpoints; and 4. no custom root certificates need to be installed by users. The interaction with the traffic is fine-grained. The use of read or write keys on specific portions of the traffic, can be used to detect changes introduced by external third parties or by middleboxes with read-only privileges. Key distribution can be done by classical IP configuration mechanisms like DHCP or DNS.

The main drawback of mcTLS is that it is modified version of TLS v1.2 that needs endpoints to implement it. In addition, [25] casts a shadow of doubt in some security aspects of mcTLS.

- ## Middlebox Secutiry Protocol

ETSI is working in its Technical Committee on Cyber-security (TC CYBER) on the Middlebox Secutiry Protocol (MSP) [26], which is derived from mcTLS. In addition to reading and modifying the stream, MSP provides for auditable insertions and deletions as well as middlebox discovery. In addition, MSP addresses vulnerabilities of mcTLS(as, e.g. [25]).

## • Middle-box TLS

Middle-box TLS (mbTLS) [27] follows the same line of mcTLS in that it provides for interaction with middleboxes. However, the approach is broader, because it considers both Network and application layer middleboxes, as well as with application-layer delegates like Content Delivery Networks (CDNs). It is immediately deployable on legacy TLS endpoints and, since it uses standard TLS v1.2 handshake with some additional extensions and records, it will traverse current firewall setups. In addition to user and out-of-band configuration, mbTLS provides for middleboxes to be discovered in-band. Regarding the handling of multi-party settings, mbTLS guarantees that the chain of middleboxes between two endpoints is visited in a specified order, hides modifications on traffic to outside observers and provides attestation facilities to endpoints regarding the Software running in middleboxes.

Regarding TLS 1.3, the authors foresee that the handshake can mbTLS can be evolved to support it with minor modifications and a potential delay in the Finished handshake of an additional round-trip.

## • EFGH

EFGH is a pluggable extension to the TLS protocol that allows middleboxes to be explicitly introduced in the client-server path without affecting the security guarantees of the end-to-end channel. The trusted middlebox is granted access to a subset of the traffic that the principals have agreed upon. The capability is negotiated between the principals through the TLS handshake extension mechanism. Upon completion of the handshake, and if both parties support EFGH, a one round-trip message exchange is required for key establishment between the client, the server and the proxy. EFGH strongly stresses the end-to-end trust relationship integrity and allows the client and server to selectively expose exchanged traffic to trusted intermediaries via a modified TLS record format.

# Path Layer UDP Substrate (PLUS)

The Path Layer UDP Substrate (PLUS) [28] follows a different approach by introducing a path layer between the IP layer and the encrypted upper layer protocols for signaling between endpoints and middleboxes while leaving the upper layer encryption unimpaired. The protocol is designed to support basic stateful in-network functionality (i.e. network state maintenance and measurement). In addition, PLUS also provides the possibility of adding extensions through so-called Path Communication Functions (PCFs) to provide a flexible and (transport-)protocol-independent mechanisms which also allows on-path devices to communicate with end-points. For information from the end-points to the network the protection context of the upper layer protocol is used to ensure integrity of information that are not meant to be alter by the network.

**PLUS and the IETF**

PLUS was presented in a Birds-of-a-Feather (BoF) session at the IETF 96 in Berlin [29]. A BoF is a meeting that brings together people to discuss a new proposal. The meeting was well-attended with many people understanding the need for this work and showing support for the approach. However, privacy and trust concerns were also being discussed, given introducing the a path layer explicitly acknowledges changes to the Internet model. The outcome of the meeting was inconclusive, but the Internet Research Task Force (IETF) has since started to explore the research issues associated with the broader topic of path aware networking.

PLUS's main merits are that it offers signals to perform basic network maintenance as well as providing functionality for interacting with middleboxes. However, as PLUS is a new separate protocol that needs to be introduced below a transport protocols, deployment depends on the adoption at the endpoint which ideally would happen in conjunction which new protocols that are under development in the IETF.

# Analysis and Recommendations

The different middlebox cooperation mechanisms presented above have different requirements with regard to the trust relationships and the coordination between the different elements in the communications path. In addition, they exhibit a different behaviour when inserted in the path and, last but not least, require different modifications to a state-of-the-art

| | Cryptographic Context | Coordination | On-path Inspection/ Modification | Deployment |
|---|---|---|---|---|
| **mcTLS** | E ↔ M ↔ E | client → middlebox → server | Both | Both endpoints need upgrade |
| **mbTLS** | E ↔ M ↔ E | No, but MB tries out server | Both | Client may be legacy |
| **Middlebox Security Protocol** | E ↔ E | E ↔ M<br>E ↔ E | Both | Changes in server only. Client is standard TLS1.3 |
| **EFGH** | E ↔ M ↔ E | both endpoints need to agree | Both | Custom header in packet frame |
| **PLUS** | E ↔ E | support on both endpoints | only explicit signaling | New protocol stack |

deployment. For an in-depth analysis of some mechanisms to expand TLS to cope with middleboxes, good work is provided by [30] (cf. Table 1, page 45).

Table 1 provides a different approach focusing on deployment challenges and the underlying coordination and cryptographic exchange mechanisms needed. First, it should be noted that all TLS-based mechanisms aim at enabling content inspection or even modifications of some or all data, while PLUS is an proposal for explicit signalling of additional information that support in-network functions. As such PLUS does require any cryptographic relationship with any middlebox, while most of the other proposals do. The TLS-based approaches different in how cooperation is negotiated and who needs to agree or will be aware of the middlebox(es). Respectively, some deployment scenarios need either both endpoints to upgrade and support the new mechanism or protocol, while in other cases only one endpoint can communicate with an enabled middlebox.

Based on this summary and on the discussion of the different approaches in Sections 3.2 and 3.1, we recommend to take the following considerations into account when designing mechanisms for encrypted traffic to interact with middleboxes:

1. end-to-end integrity of the communications must be guaranteed,

2. only strictly required information should be exposed and exclusively to the middleboxes that needs it,

3. the introduction of new features is better accepted if done in an evolutionary way (i.e. backward- compatible deployments), and

4. support mechanism needed for deployment should be minimised (i.e. avoid depending on additional configuration being managed by DHCP, DNS, etc.).

# Call for Actions

A set of seemingly conflicting trends have started to evolve in parallel. On the one hand, middleboxes have sought to provide in-network services that have been widely deployed as a fundamental component of the (IP) infrastructure, while on the other hand, the end-to-end connections using the IP infrastrcuture is increasingly encrypted for various legitimate reasons. These two trends are set to collide, because middleboxes rely upon information that becomes hidden from them when encryption is used.

Legal requirements (e.g. lawful intercept) and upcoming standardisation may force Network Operators deployment of specific middlebox cooperation mechanisms. However, a Zero-day approach has become unthinkable, and therefore a solution that can cooperate with legacy or encryption-only deployments needs to be considered.

Middlebox cooperation in encrypted end-to-end communications is a topic that has come to polarise the community. Acknowledging legitimate privacy concerns while at the same time understanding the need for mechanisms to support operation of the Internet needs to pave the way for an informed discussion among stakeholders in a constructive environment and without prejudices. With the precedent of the successful MAMI Management and Measurement Summit (M3S) and taking the example of the Mutually Agreed Norms for Routing Security (MANRS) initiative [31] fostered by the Internet Society (ISOC), there is a need for a meeting point where actors can discuss and agree on basic principles, such as the exposed information and its semantics for middlebox cooperation. A one-sided approach from any organisation or interest group is expected to be counterproductive.

...

This work is partially supported by the European Commission under Horizon 2020 grant agreement no. 688421 Measurement and Architecture for a Middleboxed Internet (MAMI), and by the Swiss State Secretariat for Education, Research, and Innovation under contract no. 15.0268. This support does not imply endorsement.